# Artificial Intelligence in Achieving Sustainable Development Goals

Hoe-Han Goh[1]*

[1]Institute of Systems Biology, Universiti Kebangsaan Malaysia, Bangi, Selangor, 43600, Malaysia.

*Corresponding author. Email: gohhh@ukm.edu.my

**Abstract:** This perspective illustrates some of the AI applications that can accelerate the achievement of SDGs and also highlights some of the considerations that could hinder the efforts towards them. This emphasizes the importance of establishing standard AI guidelines and regulations for the beneficial applications of AI.

**One-Sentence Summary:** AI applications can empower humanity in saving our planet and us.

The advent of information and communications technology (ICT) and Internet-of-Things (IoT) has culminated in the fourth industrial revolution (4IR) and ushered in the big data era of artificial intelligence (AI)-enabled precision and personalization via the convergence of physical, biological, and digital technology. AI is game-changing by maximizing objective functions for the optimal outcome.

In agriculture, AI allows many labor-intensive processes to be automated (*1*), including autonomous systems in transport, logistics, and supply chain. Smart systems can be deployed for farm monitoring, management, husbandry, operation, and surveillance as well as implementing smart contracts with blockchain for automated transactions. They have the advantages of transparency, reliability, efficiency, and cost-saving. Speed breeding of crops is made possible with smart greenhouses regulating the photoperiods via LED lightings to shorten the time to flowering and fruiting. Automated vertical farms integrated with aquaculture have been adapted for urban farming to reduce the carbon footprint of transporting fresh produce. Farming on the rooftop of office buildings and skyscrapers not only maximizes the space usage but also adds greenery to the urban landscape against pollutions while harvesting the solar energy for food production and generating renewable electricity with solar panels. Soil-less plantation methods such as aquaponics or aeroponics are no longer restricted to vegetables but also tuberous crops like potatoes. High-tech sustainable farming with waste-water treatment for zero waste is in line with the circular economy and can even prolong crop shelf-life.

AI-enabled precision farming integrates climate sensors with automated irrigation and fertilizing systems can function efficiently with minimal labor. The digital twin of greenhouses permits real-time monitoring and controls. Smart sweepers or harvesters are made possible with continuous machine learning (ML) based on virtual training with augmented reality (AR) for the robots to recognize fruits of different ripening stages from hard-to-reach areas. For large-scale farming, IoT-enabled monitoring of soil and weather conditions permits the automation of farm husbandry. Diverse robotic and drone technologies are available for weeding, fertilizing, seeding, pruning, and harvest in the field. Satellite and drone imaging allow field assessment to inform agrichemical applications and measure crop productivity for yield prediction. Governmental Agri-Food Transformation Fund will encourage adoption of precision farming practices worldwide. Active partnerships of colleges with local farms can be forged via transformative project-based internships. Beyond that, workforce upskilling initiatives are timely to be deployed via Technical and Vocational Education and Training (TVET) and micro-credentials for all the sectors and industries. AI can contribute to personalized learning (*2*) for quality education.

While developing technology to increase food production with efficient farming, the green citizenry should also be cultivated through awareness campaigns to reduce consumptions and food wastes. AI can help reducing wastes with a more efficient supply chain (*3*). Automated food delivery to predict daily orders based on consumer big data for ingredient logistics and food waste minimization promotes a more sustainable food system. The impacts will be tremendous when applied to large fast-food chains around the world to optimize resources. Another transformative change is to shift from animal-based to plant-based proteins that will greatly cut the emissions of greenhouse gases (GHG). To change dietary preference, a Chilean company, NotCo, uses an AI algorithm "Giuseppe" to reproduce animal taste and texture based on molecular flavor profiles from hundreds of thousands of plant ingredients to look, cook, blend, and taste just like the animal produce. Such a food-tech disruptor not only can save the environment but also ensure food security for the future.

Agriculture is a critical component of national security and world health because food shortages will lead to famine and political unrest. At Descartes Labs, physicists use the principles of physics and light with remote sensing to predict when disease, disaster, or war might strike based on historical satellite imagery. The similarity search engine allows object recognition for monitoring land usage. This allows nationwide prediction of crop yield in the US cornfields and Asian rice paddies with crop health inferred from satellite infrared bands. Weather forecast can also better prepare farmers with planning to minimize losses from possible drought or flood based on water availability models. Earth observation data can also help us to understand socioeconomic and environmental trends to predict how today's decisions will impact tomorrow's society. It is possible to infer poverty and food production situation in the Middle East and North Africa for pre-emptive food aids in times of crop failure to prevent famines months before it happens, especially in troubled areas. In case of crisis, every nation needs a future-ready food hub for global coordination in assessing and regulating food security and safety to ensure reliable sources of food supplies and accountability within the intertwined global food supply chain.

AI is also useful for environmental conservations to save wildlife, biodiversity, and the world to halt the sixth mass extinction. For example, Intel's TrailGuard that is equipped with a vision processing unit (VPU) can detect a human from body shapes, facial geometry, gait, and movement from any angle and light condition via intelligent image recognition. It serves as a hidden park ranger that works 24/7 in tracking wildlife poachers with an automated alert system to the control center to prevent poaching before it happens. The current predictive models come with unprecedented precision through continuous data training. The models are ever-improving without a priori settings for unsupervised analysis. AI can identify patterns in better forecasting disasters or calamities to prevent catastrophe. ShakeAlert® applies ML to discern noises from signals of earthquakes via hundreds of vibrational sensors scattered around big cities like San Francisco and Seattle.

Despite that AI technology can be an accelerator for achieving the United Nations Sustainable Development Goals (UN SDGs), we must realize that AI is not a solve-it-all. AI solution is often an evolutionary process instead of a turnkey solution. Transformative solutions are much rarer than incremental solutions. Passionate purpose-driven leaders are essential to champion sustainability as the main goal in organizations, society, local governments, and countries. In this respect, leaders and industry players from different sectors play important roles in the future of AI. Outlook and perspectives on AI involve active discussions from all stakeholders on the issues, considerations, and implications of different AI applications (Fig. 1). AI could play both enabler and inhibitor roles in achieving the SDGs (*4*). Hence, holistic assessments will be needed for each AI application by considering the societal, economic, and environmental impacts of digital sustainability (digitainability) (*5*).

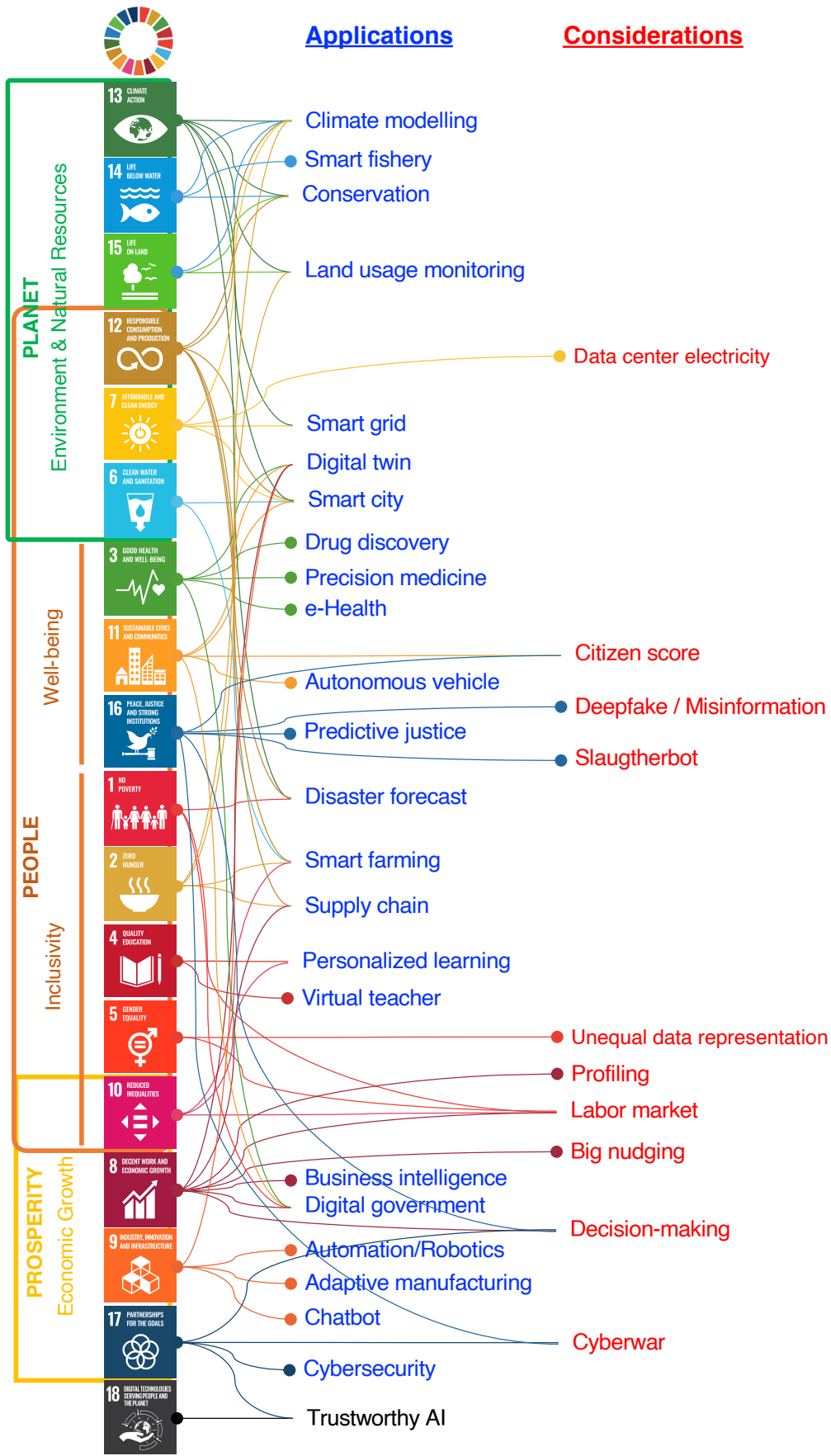

**Fig. 1. Applications and considerations of AI in relation to the 17 Sustainable Development Goals.** The 18[th] SDG of Digital Technologies Serving People and the Planet has been proposed to ensure the digital age supports people, planet, prosperity, peace, and partnerships. The arrangement and grouping of SDGs into planet, people, and prosperity has not been reviewed by the United Nations and does not reflect its views. The applications and considerations are not an exhaustive list to reflect digital interdependence. Some connections between the applications/considerations and SDGs are not shown for simplicity as represented by the rounded ends.

Some of the pressing issues and considerations include transparency, equality, automated decision-making, and ethical standards of profiling (*6*). The infrastructural vulnerability has shifted from the millennial digital divide to digital interdependence. It is inevitable for an accelerated trend of digitalization and AI applications to achieve SDG indicators. Therefore, practical implications of AI-based technological deployment need to be considered beyond sectorial silos to reflect the cross-cutting and interconnectedness of SDGs (Fig. 1). For instance, the digital twin can be represented in multiple SDGs, while the electricity demands of ICT and data centers will need solutions from renewable energy sectors. This requires a systems approach to consider all interactions among the targets of SDGs (*7*). The current speed of technological advancements surpasses individual adoption and governmental regulations of the double-edged AI. To reap the maximum potential from doing good with AI such as smart fishery (*8*), smart grid (*9*), smart city (*10*), and e-health (*11*), we must impose controls on "big nudging", citizen score, and deepfake; while balancing societal polarization with ethics. Gender inequality in data with male-dominated datasets and manpower should be clarified to avoid discrimination and bias in AI applications (*12*).

Internet history leaves a trail of digital exhaust that is hoovered for behavioral prediction and micro-targeting with the possibility of becoming a social contagion, such is the case of election manipulation by Cambridge Analytica via Facebook. Such a scenario is a reminder to government policymakers and regulators to think twice about the aggressive lobbying of tech giants against regulations. Harmful misuses must be prevented via international dual-use aware policy and regulations such as that of the Biological Weapons Convention. Countermeasures should be implemented against AI terrorism such as cyber-attacks, slaughterbots, killer drones, hacks, glitches, exploitations, and manipulations. The 23 Asilomar AI principles (*13*) is a good start for global shared AI visions of equitable benefits and prosperity for all without compromising people's rights, privacy, autonomy, dignity, and freedom. EU's current guidelines on trustworthy AI can be adapted for other regions. The guidelines and good practices of ethical AI in a human context (*14*) have been advocated with direct impacts on people's life and future. AI practitioners should consider the long-term implications and impacts for AI applications and data-driven innovations to be aligned with SDGs, such as the AI-enabled climate actions. AI applications should be ethics-driven, context-aware, and inclusive for use responsibly. Globally, the field needs intergovernmental regulatory harmonization and open policy for data transportability. We need shared visions and wisdom to keep AI beneficial to mankind (*15*) by making sure the machines or robots understand, adopt, and retain humanity's goals. An additional SDG of Digital Technologies Serving People and the Planet (SDG18) has been advocated by the Montreal Statement on Sustainability in the Digital Age (*16*). This is to ensure the digital age supports people, planet, prosperity, peace, and partnerships.

    Robotics and AI have been applied in bionics for augmenting and enhancing human capability. This is especially meaningful for handicapped people to acquire artificial limbs and regain sensory perception. On the other hand, it remains challenging to create a robot with human-like dexterity and mobility for common tasks that we often take for granted, such as grabbing a drink from the fridge. This requires visual perception through computer vision in an unstructured environment that the robot will take time to process and learn. This is because a robot can only "see" binary numbers, unlike human perception. The good news is that once computer intelligence is achieved, it can be easily replicated through cloud deployment to all machines or robots.

Although we are still far from sci-fi humanoids like the terminator, many AI scientists are striving for a person-like robotic system with a body and mind analogous to a human being that can pass the Turing test. To achieve this, advancements have been made in chatbot, gaze tracking, object detection, emotion detection, and copy of human sensorium. Humanoids are ever-improving with advanced materials like carbon fiber skeleton and Kevlar muscle equipped with pneumatic actuators using compressed air that is analogous to human musculoskeletal system with proprioceptors. The ultimate goal is to create a virtual mind of a humanoid beyond knowledge, skills, and language towards an ultimate conscious machine that thinks on its own. Consciousness is related to awareness and a sense of being. This is how we fit in time and space with sensory memories that shape our manners. "Synth" is an artificial recreation of memories from texts via natural language processing (NLP), which can provide backstories for the creation of new memories in a humanoid. Cracking the neural code in human brain science will have profound implications to the mind of a synth as revolutionary as deciphering the genetic code for biology.

In the process of advancing humanoid, AI experts appreciate more about humanity in terms of intuition, creativity, and empathy that are lacking in AI. Human-level common sense poses the greatest challenge in AI programming. In this regard, AI transformation should be considered complementary instead of total replacement to human society. Reaching the technological singularity was once a fantasy. In the future, we might succeed in creating digital intelligent life on Earth while SETI is searching for extraterrestrial intelligence via Allen Telescope Array (ATA) using AI in teaching machines to learn from mining vast volumes of data from the universe and looking for the exceptions to the norm. A synthetic superintelligence will transcend biological intelligence, resulting in unforeseeable changes to human civilization. Will this be a new dawn of human society and moral principles? Who will be responsible for robotic crimes? Will there be robot rights when they can be considered their own being with volitions? How about synth-ethics?

By then, will we be pondering on the existential questions of humanity if we can upload our minds into the digital self?

Meanwhile, AI technology is already rapidly changing the world of how we live and work. AI can create new opportunities and potentially filling the gaps of works that no one wants to do. Yet, the labor market has been greatly impacted by reduced dependency on laborer, resulting in the unemployment of the low-skill workers and widening the gap of the society that need urgent attention in many countries (*10*). Not all countries can afford to provide the social safety net for the foreseeable income inequality between the AI-technology haves and have-nots. This raises a utopian proposal of a global welfare state under the positive vision of shared prosperity to avoid destructive conflicts or wars between countries. Hopefully, AI-based technology can help to save the planet and ourselves for the betterment of humanity before the tipping point of global destruction.

REFERENCES AND NOTES

1. P. Zhang *et al.*, Nanotechnology and artificial intelligence to enable sustainable and precision agriculture. *Nature Plants* **7**, 864–876 (2021).
2. A. Bozkurt, A. Karadeniz, D. Baneres, A. E. Guerrero-Roldán, M. E. Rodríguez, Artificial intelligence and reflections from educational landscape: A review of AI studies in half a century. *Sustainability (Switzerland)* **13**, 1-16 (2021).
3. A. Di Vaio, F. Boccia, L. Landriani, R. Palladino, Artificial intelligence in the agri-food system: Rethinking sustainable business models in the COVID-19 scenario. *Sustainability (Switzerland)* **12**, (2020).
4. R. Vinuesa *et al.*, The role of artificial intelligence in achieving the Sustainable Development Goals. *Nature Communications* **11**, 233 (2020).
5. S. Gupta, M. Motlagh, J. Rhyner, The digitalization sustainability matrix: A participatory research tool for investigating digitainability. *Sustainability (Switzerland)* **12**, 1-27 (2020).
6. T. Hagendorff, The ethics of AI ethics: An evaluation of guidelines. *Minds and Machines* **30**, 99-120 (2020).
7. D. Le Blanc, Towards integration at last? The Sustainable Development Goals as a network of targets. *Sustainable Development* **23**, 176-187 (2015).
8. S. H. Ebrahimi, M. Ossewaarde, A. Need, Smart fishery: A systematic review and research agenda for sustainable fisheries in the age of AI. *Sustainability (Switzerland)* **13**, (2021).
9. M. Massaoudi, H. Abu-Rub, S. S. Refaat, I. Chihi, F. S. Oueslati, Deep learning in smart grid technology: A review of recent advancements and future prospects. *IEEE Access* **9**, 54558-54578 (2021).
10. O. Pilipczuk, Sustainable smart cities and energy management: The labor market perspective. *Energies* **13**, (2020).
11. N. Schwalbe, B. Wahl, Artificial intelligence and the future of global health. *The Lancet* **395**, 1579-1586 (2020).

12. D. Cirillo *et al.*, Sex and gender differences and biases in artificial intelligence for biomedicine and healthcare. *npj Digital Medicine* **3**, (2020).
13. https://futureoflife.org/ai-principles/
14. J. Salgado-Criado, C. Fernandez-Aller, A wide human-rights approach to artificial intelligence regulation in Europe. *IEEE Technology and Society Magazine* **40**, 55-65 (2021).
15. J. Cowls, A. Tsamados, M. Taddeo, L. Floridi, A definition, benchmark and database of AI for social good initiatives. *Nature Machine Intelligence* **3**, 111-115 (2021).
16. https://sustainabilitydigitalage.org/montreal-statement/

ACKNOWLEDGMENTS

H.-H.G. acknowledges funding supports from the Ministry of Higher Education Malaysia (FRGS) and UKM Research University Grant (DIP).